\documentclass[draft]{agujournal2019}
\usepackage{url}
\usepackage{lineno}
\usepackage{soul}
\usepackage{amsmath}
\usepackage{amssymb}

\drafttrue

\journalname{Space Weather}

\begin{document}

\title{Beyond Mean Solar Wind Conditions: Turbulence-Aware Forecasting of the AE Index}

\authors{Cara L. Waters\affil{1}, Christopher H. K. Chen\affil{1}, Mathew J. Owens\affil{2}}

\affiliation{1}{Department of Physics \& Astronomy, Queen Mary University of London, London, UK}
\affiliation{2}{Department of Meteorology, University of Reading, Reading, UK}

\correspondingauthor{Cara L. Waters}{c.waters@qmul.ac.uk}

\begin{keypoints}
\item Including solar wind turbulence parameters improves short-term forecasts of AE index compared to models using average conditions alone.
\item Improved forecasts reduce false alarms and better capture extreme geomagnetic events, supporting more reliable operational decision making.
\item Turbulence-aware forecasts provide insight into how solar wind variability influences energy transfer in Earth’s space environment.
\end{keypoints}

\begin{abstract}
The auroral electrojet (AE) index is a key indicator of high latitude geomagnetic activity and is widely used in operational space weather monitoring, yet forecasting AE from upstream solar wind conditions remains challenging due to nonlinear coupling, internal magnetospheric dynamics, and multiscale variability. We test whether incorporating solar wind turbulence improves short timescale AE forecasts beyond models based only on mean solar wind and interplanetary magnetic field parameters. Two gradient boosted decision tree (XGBoost) models are developed using near-Earth solar wind observations: a baseline model using standard mean parameters and a turbulence-aware model that additionally includes measures of fluctuation amplitude, intermittency, and Alfv\'enic structure. Both models achieve peak performance at short lead times, with correlations exceeding 0.8 at 60 minutes. However, while the baseline model exhibits a clear skill peak at 75 minutes, the turbulence-aware model maintains comparable skill across 60--90 minute horizons, indicating reduced degradation with lead time. The turbulence-aware model also provides consistent improvements over both the baseline and persistence and, critically, improves forecast robustness for high-impact events. Cost–loss analysis shows that, for the baseline model, economic value decreases systematically with increasing AE threshold and the range of cost–loss ratios yielding positive value narrows. In contrast, the turbulence-aware model maintains an approximately constant zero-value cost–loss threshold across all event levels, indicating stable economic usefulness even for extreme AE conditions. This demonstrates that turbulence provides complementary, scale-dependent information beyond mean solar wind parameters, improving both forecast performance and decision-relevant value for operational space weather applications.
\end{abstract}

\section*{Plain Language Summary}
The auroral electrojet (AE) index is a measure of electrical currents flowing in Earth’s upper atmosphere near the poles and is widely used to monitor space weather conditions that can affect satellites, power systems, and radio communications. Forecasting AE is challenging because Earth’s space environment depends not only on the average conditions of the solar wind --- a stream of charged particles from the Sun --- but also on how turbulent and variable that solar wind is. In this study, we investigated whether including information about solar wind turbulence improves short-term forecasts of AE. We compared two machine learning models: one that uses only average solar wind and magnetic field conditions, and another that also includes simple measures describing solar wind turbulence. We found that the model including turbulence performs better overall, especially during periods of strong geomagnetic activity. Importantly, it is better at forecasting extreme AE events and produces fewer false alarms, which is important for space weather forecasting and decision making. These results show that accounting for solar wind turbulence can improve the reliability of geomagnetic activity forecasts without making prediction models more complex, helping users respond more effectively to space weather hazards.

\section{Introduction}

The auroral electrojet (AE) index is a widely used proxy for high latitude geomagnetic activity and substorm-related ionospheric current systems. It is constructed from the range of the horizontal magnetic perturbation (east to west) observed at a network of auroral-zone ground magnetometers, providing a global measure of electrojet intensity and its variability on minute to hour timescales \cite{davis1966, mayaud1980}. Because AE responds rapidly and strongly to changes in solar wind driving and reflects the efficiency of magnetosphere--ionosphere coupling, it has long been used as a diagnostic of energy transfer in the geospace system and as an indicator of disturbed near-Earth space conditions for operational forecasting.

Forecasting AE from upstream solar wind conditions is challenging for two closely related reasons. First, the coupling between the solar wind and magnetosphere is highly nonlinear and depends on multiple interacting drivers, including the interplanetary magnetic field (IMF) orientation and strength, solar wind speed, density, and dynamic pressure. At the most basic level, southward IMF enables magnetic reconnection at the dayside magnetopause, opening flux and driving large-scale convection in the Dungey cycle framework \cite{dungey1961, kan1979}. This conceptual picture underpins many empirical descriptions of solar wind--magnetosphere coupling and has motivated the development of a wide range of empirical coupling functions and energy input proxies designed to condense the geoeffective solar wind driving into a small number of physically motivated parameters \cite<e.g. >{perreault1978, newell2007}. While such functions capture broad statistical relationships, they necessarily simplify the underlying dynamics and do not uniquely represent the pathways by which energy is transferred, stored, and released within the coupled magnetosphere--ionosphere system. Comparative studies have further shown that the performance of coupling functions depends on activity level, timescale, and the specific geomagnetic response under consideration \cite{finch2007, lockwood2022}.

Second, geomagnetic activity is not controlled solely by slowly varying mean solar wind conditions. Instead, it is also sensitive to fluctuations, intermittency, and multiscale structure in the solar wind and IMF, which can modulate reconnection rates, plasma transport, and energy conversion processes across a wide range of temporal and spatial scales \cite{borovsky2003, jankovicova2008, lyons2003}. As a result, two intervals with similar mean solar wind properties can produce different geomagnetic responses, highlighting the importance of variability and turbulence in solar wind--magnetosphere coupling. This variability is reflected not only in the solar wind but also in AE itself, which exhibits intermittent, scale-dependent behaviour and signatures of both directly driven and internally generated dynamics \cite{hnat2003, consolini2005, freeman2000}.

The potential importance of solar wind turbulence in driving magnetospheric dynamics has been recognised since early theoretical work. Initial hypotheses proposed that solar wind--magnetosphere interaction could be described in terms of a viscous-like coupling, in which turbulent interactions at the magnetopause transfer momentum and energy into the magnetosphere, driving large-scale convection \cite{axford1961}. Order-of-magnitude estimates suggested that such viscous energy input could be comparable to the energy dissipation associated with geomagnetic storms, and this mechanism was invoked to explain features such as the extended length of the magnetotail and asymmetries in particle trapping boundaries \cite{axford1964}.

Subsequent observational and numerical studies have shown that solar wind plasma transport across the magnetopause can occur through multiple boundary processes, including Kelvin-Helmholtz instability and associated vortex-induced mixing along the flanks, which can facilitate the entry of magnetosheath plasma even under northward IMF conditions \cite{hasegawa2004, nykyri2001}. In addition, turbulence generated upstream of the bow shock and within the magnetosheath can modify the structure and variability of the magnetopause, influencing the efficiency and intermittency of energy transfer into the magnetosphere \cite<e.g. >{zimbardo2010}. In this modern view, viscous-like interaction and turbulence-mediated transport are understood as contributing pathways within a broader coupling system.

Observational studies have since provided further evidence linking solar wind fluctuations to geomagnetic activity. $B_z$ fluctuations at low frequencies have been shown to impact geomagnetic activity strongly during both strongly northward and southward IMF \cite{osmane2015}. Alfv\'enic turbulence in fast solar wind streams has been shown to act as an effective driver of auroral activity, particularly during solar minimum, even under relatively weak mean solar wind forcing \cite{damicis2009}. During solar maximum, when solar wind fluctuations are more strongly influenced by advected coronal structures, this relationship weakens, suggesting a change in the dominant coupling regime \cite{damicis2010}. More generally, statistical analyses of solar wind fluctuations have demonstrated that while skewness remains close to zero across scales and does not strongly distinguish between quiet and active intervals, kurtosis increases significantly at smaller scales and during disturbed periods, reflecting enhanced intermittency \cite{jankovicova2008}. This supports the idea that non-Gaussian, intermittent fluctuations may modulate the efficiency of solar wind--magnetosphere coupling, beyond what is captured by mean solar wind parameters alone \cite{borovsky2003}. Measures characterising the Alfv\'enic nature of the solar wind, including cross helicity and residual energy, have also been shown to vary systematically with geomagnetic activity and solar cycle phase, further supporting a link between turbulence properties and geoeffectiveness \cite{damicis2007, damicis2011}. A recent review has reinforced this overall picture, arguing that turbulence properties in the upstream solar wind can contribute significantly to geomagnetic response, especially at high latitudes \cite{damicis2020}.

Despite decades of study, solar wind--magnetosphere coupling remains incompletely understood. Several key physical processes are still insufficiently characterised for predictive modelling, including the detailed control of reconnection, plasma entry and mass loading pathways, and nonlinear feedbacks and saturation effects within the coupled system \cite{borovsky2021}. These limitations pose a challenge for space weather forecasting, particularly for quantities such as AE that respond rapidly to both large-scale driving and smaller-scale variability.

In this context, data-driven approaches provide a complementary avenue for improving geomagnetic activity forecasts. Modern machine learning methods can represent nonlinear dependencies without prescribing a fixed functional form and can integrate information from a large number of correlated input variables. A substantial body of work has therefore explored data-driven approaches to AE prediction, including early feed-forward and recurrent neural networks, dynamic neural networks, and nonlinear system identification techniques \cite{takalo1997, weigel1999, gleisner2001, pallocchia2008, gu2019}. These studies demonstrate that meaningful predictive skill can be obtained from L1/near-Earth solar wind measurements, but they also highlight persistent limitations: forecast skill can vary strongly between quiet and active intervals, and models trained primarily on mean or coarsely average solar wind parameters often struggle to capture rapid intensifications and sharp gradients. More broadly, machine learning methods are increasingly central to space weather nowcasting and forecasting, offering flexible nonlinear function approximation while raising new questions about robustness, generalisation, and physical interpretability \cite{camporeale2019}.

Previous studies of AE forecasting span several generations of modelling approaches and provide quantitative context for evaluating current performance. Early empirical, linear, and neural network-based models demonstrated that statistically significant prediction skill is achievable, particularly at short lead times, but also revealed clear limitations. Dynamic and recurrent neural network models driven by upstream solar wind parameters typically achieved correlation coefficients of order 0.6--0.7 for lead times of tens of minutes to around one hour, with systematic underestimation of peak AE values and reduced skill during highly disturbed intervals \cite{takalo1997, weigel1999, gleisner2001}. Empirical solar wind-driven models have reported higher correlations at short lead times, with correlations approaching 0.85--0.9 for 10 minute averaged AE and root mean square errors of order $\sim$100 nT, though performance degrades as geomagnetic activity increases \cite{luo2013}. More recent data-driven and system identification approaches have improved overall skill at hour-scale lead times, with correlations commonly reported in the range of $\sim$0.85--0.88 and prediction efficiencies of $\sim$0.7--0.75, while still exhibiting increasing error and bias during extreme events \cite{gu2019}. Deep learning models developed in the past few years report comparable headline correlations for similar lead times, although error metrics are often reported for normalised quantities, complicating direct comparison with earlier studies, but can achieve correlations of greater than 0.85 \cite{zou2024}. Taken together, these results suggest that, for solar wind-driven AE prediction, forecast skill has improved incrementally over time but appears to approach an upper bound set by internal magnetospheric dynamics, nonlinear coupling processes, and the influence of multiscale variability \cite{bargatze1985, camporeale2019}. This motivates continued investigation into whether additional physically meaningful descriptors of the upstream solar wind --- beyond mean parameters alone --- can provide further gains in forecast accuracy and robustness.

Comparable efforts have targeted other electrojet and auroral indices, reflecting both the diversity of available ground-based products and differing operational needs. Forecasting of the westward electrojet (AL) index has been addressed using neural network approaches \cite{amariutei2012}, while more recent work has applied machine learning to the SuperMAG SML index \cite{maimaiti2019}, including using an imbalanced regression neural network \cite{chu2025}, which predicts SML from solar wind inputs without relying on past index values and is specifically designed to recover strong to extreme events that conventional mean squared error training tends to underestimate. Regional indices derived from dense magnetometer networks, such as the IMAGE IL/IU/IE indices, have likewise been used to capture the spatio-temporal structure of high latitude activity \cite<e.g. >{tanskanen2009, ventriglia2025}. Across these indices, however, the upstream solar wind is typically represented through mean parameters or empirical coupling functions; the explicit turbulent nature of the driving flow is rarely incorporated as a predictor.

In this study we test the hypothesis that explicitly including turbulence information in the upstream drivers improves short-timescale AE forecasts beyond what is achievable with mean solar wind parameters alone. We build two gradient boosted decision tree models using XGBoost \cite{xgboost} to forecast AE from time-shifted near-Earth solar wind conditions. The first model uses a standard set of mean solar wind and IMF predictors (e.g. $V$, $n$, $|\mathbf{B}|$, IMF components). The second model augments these predictors with explicit turbulence descriptors computed from higher-cadence fluctuations over rolling windows, intended to capture the amplitude, intermittency, and Alfv\'enic structure of the solar wind (e.g. fluctuation levels with $\delta \mathbf{B}$ and $\delta \mathbf{V}$, and dimensionless measures motivated by MHD turbulence such as cross helicity and residual energy). Such quantities have been argued to modulate geoeffectivity and coupling efficiency even under comparable mean driving by altering the occurrence and effectiveness of reconnection and by injecting variability across magnetospheric response times \cite{borovsky2003, jankovicova2008, damicis2020}. By comparing these models, we determine the added value of turbulence-aware models for AE prediction.

Section \ref{sec:data} describes the dataset used for this study. The methods, both construction of the machine learning model and the parameters used, are described in Section \ref{sec:methods}, along with the evaluation methods. Section \ref{sec:performance} uses these evaluation methods to determine the additional forecasting skill provided by a turbulence-aware model, and Section \ref{sec:parameters} uses interpretation methods to determine the physical drivers of this. Finally, Section \ref{sec:impact} discusses the benefit for forecast users, Section \ref{sec:discussion} discusses these findings and future paths for space weather forecasts, and Section \ref{sec:conclusions} concludes.

\section{Data} \label{sec:data}

\begin{figure}
    \centering
    \includegraphics[width=\linewidth]{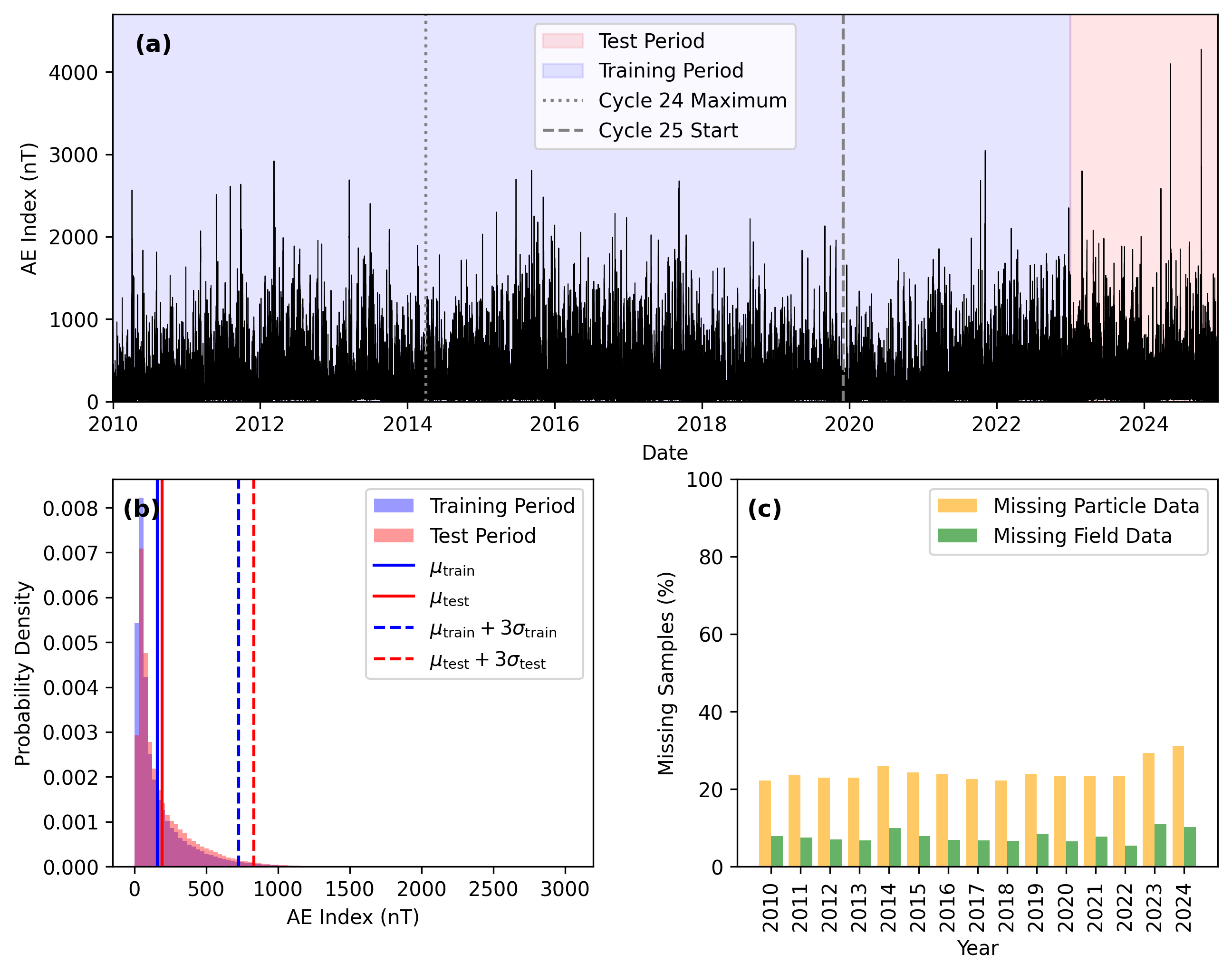}
    \caption{An overview of the period of OMNI data and AE index used, with (a) the AE index over the entire period, with the training period highlighted in blue, the test period highlighted in red, and the maximum of solar cycle 24 and start of solar cycle 25 indicated by vertical lines, (b) histograms of AE index over both test periods, with means and means plus three standard deviations indicated, and (c) the percentage of missing data for both field and particle data per year.}
    \label{fig:omni}
\end{figure}

The dataset used spans January 2010 to December 2024, covering the majority of solar cycle 24 and the rising phase of solar cycle 25 (Figure \ref{fig:omni}a). This interval includes extended quiet periods and multiple episodes of extreme geomagnetic activity, providing a wide range of solar wind driving conditions.

Solar wind plasma and interplanetary magnetic field data are taken from the 1-minute OMNI dataset \cite{omni}, which combines measurements from multiple upstream spacecraft and propagates them to the Earth's bow shock nose. This propagation introduces a timing uncertainty, typically of order a few minutes, arising from the assumptions inherent in the phase front propagation technique \cite{king2005, weimer2003, case2012}. This uncertainty is small compared with the timescales over which the fluctuations we examine evolve and the forecast lead time, and we therefore do not expect it to materially affect our results. We note that, because the prediction is keyed to conditions at the bow shock while the driving solar wind is observed upstream, there is a genuine lead time between input and predicted response. In this sense this constitutes short term forecasting rather than nowcasting, as the propagation delay represents a real predictive horizon. The variables used are the GSM IMF components ($B_x$, $B_y$, $B_z$), magnetic field strength $|\mathbf{B}|$, GSE solar wind velocity components ($V_x$, $V_y$, $V_z$), and proton number density $n$.

Geomagnetic activity is characterised using the 1-minute AE index, a widely used proxy for auroral electrojet intensity and substorm activity \cite{davis1966}. The AE distribution is strongly non-Gaussian with a pronounced high-activity tail (Figure \ref{fig:omni}b), motivating conditional evaluation later in the analysis. It is worth noting that AE index is derived only from northern hemisphere data, and that asymmetries in the geomagnetic field mean the southern hemisphere response may not be identical.

The data are split chronologically to ensure temporal independence, with January 2010--December 2022 used for training and January 2023--December 2024 used for testing. This allows evaluation under previously unseen conditions during the rising phase of solar cycle 25 after training on a full solar cycle. Magnetic field data coverage exceeds 90\% over the full interval, while plasma moments exhibit larger data gaps (Figure \ref{fig:omni}c). Short gaps of up to three minutes are forward-interpolated to preserved short timescale continuity, while longer gaps remain missing. No interpolation is applied to the AE index, which is used only where observed.

\section{Methods} \label{sec:methods}

\subsection{Forecast formulation and feature construction}

The AE index is treated as the prediction target, with forecasts defined at fixed lead times $\tau \in \{30, 45, 60, 75, 90, 120, 180\}$ minutes. All input features at time $t$ are used to predict $AE(t + \tau)$, ensuring a strictly causal formulation.

To capture the multiscale nature of solar wind driving, rolling statistics are computed over multiple window lengths. For bulk solar wind parameters, rolling means are calculated over windows of 5--120 minutes. For turbulence diagnostics, statistics are computed over windows of 5--60 minutes. A minimum data availability criterion of 50\% within each window is imposed to avoid spurious values due to sparse sampling.

Turbulence features are derived from fluctuations defined relative to the rolling mean background. From these fluctuations, RMS amplitudes, higher order moments (skewness and kurtosis), and physically motivated turbulence parameters are computed. These include the normalised cross helicity and residual energy \cite<e.g. >{chen2016, schekochihin2022},

\begin{equation}
    \sigma_c(W) = 
    \frac{2 \langle \delta \mathbf{v} \cdot \delta \mathbf{b} \rangle_W}
    {\langle |\delta \mathbf{v}|^2 \rangle_W + \langle |\delta \mathbf{b}|^2 \rangle_W},
\end{equation}
\begin{equation}
    \sigma_r(W) = 
    \frac{\langle |\delta \mathbf{v}|^2 \rangle_W - \langle |\delta \mathbf{b}|^2 \rangle_W}
    {\langle |\delta \mathbf{v}|^2 \rangle_W + \langle |\delta \mathbf{b}|^2 \rangle_W},
\end{equation}
which both make use of the magnetic field normalised by the local Alfv\'en speed,
\begin{equation}
    \mathbf{b}(t) = \frac{\mathbf{B}(t)}{\sqrt{\mu_0 m_p n(t)}},
\end{equation}
where $n$ is the proton number density, $m_p$ is the proton mass, and $\mu_0$ is the permeability of free space. $\sigma_c$ of $\pm 1$ indicates unidirectional Alfv\'enic fluctuations, whereas $0$ indicates balanced turbulence, and $\sigma_r < 0$ indicates magnetically dominated turbulence, whereas $\sigma_r > 0$ indicates kinetically dominated turbulence. The final parameter derived is the magnetic compressibility,

\begin{equation}
    C_B(W) = 
    \frac{\langle \delta |\mathbf{B}|^2 \rangle_W}
    {\langle \delta B_x^2 + \delta B_y^2 + \delta B_z^2 \rangle_W},
\end{equation}
which quantifies the fraction of magnetic energy fluctuation in compressive modes. Together, these quantities describe the amplitude, intermittency, geometry, and energy balance of solar wind turbulence.

\subsection{Forecast models}

Forecasts are produced using gradient-boosted decision trees implemented with XGBoost \cite{xgboost}. In gradient boosting, an ensemble model is constructed by sequentially adding many shallow decision trees, where each new tree is trained to reduce the prediction errors (residuals) of the existing ensemble. This stage-wise optimisation corresponds to performing gradient descent in function space on a specified loss function, here the squared error objective. The final forecast is obtained as the weighted sum of the predictions from all trees, allowing the model to represent complex nonlinear relationships and interactions between input variables.

XGBoost (Extreme Gradient Boosting) is an efficient and regularised implementation of this approach. It incorporates shrinkage (learning rate control), tree depth constraints, and explicit penalty terms on model complexity to reduce overfitting, while enabling fast parallel training on large datasets. These properties make gradient-boosted trees well suited to space weather forecasting problems, where the solar wind–magnetosphere coupling exhibits nonlinear responses, threshold behaviour, and interactions across multiple temporal scales. In addition, tree-based models naturally accommodate features with different units and dynamic ranges without requiring strong distributional assumptions, and permit post-hoc interpretation through feature attribution methods.

These properties come with limitations worth noting. Tree ensembles cannot extrapolate beyond the range of target values present in the training data, since each prediction is an average over training samples within a leaf; the rarest and most extreme AE intensifications are therefore prone to systematic under-prediction. Gradient-boosted trees also have no intrinsic representation of temporal structure and treat each sample independently, so any sequential information must be provided explicitly through engineered predictors, such as the rolling window diagnostics used in this study.

Models are trained using a squared-error objective with regularisation to limit overfitting. Three model configurations are considered:

\begin{enumerate}
    \item a baseline model using only large scale solar wind parameters and AE history;
    \item a turbulence-aware model that augments the baseline with solar wind turbulence diagnostics; and
    \item a noise-control model in which turbulence features are replaced by statistically matched random surrogates, to isolate the contribution of providing a greater number of features alone.
\end{enumerate}

\subsection{Training, validation, and optimisation}

All models are trained using a strictly time-ordered procedure to reflect the real forecasting setting. In particular, data from later times are never used to inform predictions at earlier times. This prevents ``information leakage'', in which knowledge of the future could artificially improve model performance.

Model settings (hyperparameters), such as the complexity of the decision trees and the rate at which the model learns from new information, are automatically tuned using the Optuna optimisation framework \cite{optuna}. To do this, the training interval is divided into two parts: an initial portion used to fit the model, and a final 20\% subset used for validation. Model performance on this validation subset provides an objective way to select hyperparameters that generalise well to unseen data.

During training, early stopping is applied. This means that model fitting is halted if performance on the validation data no longer improves, helping to avoid overfitting, where the model learns noise or event-specific details rather than robust physical relationships.

After the optimal hyperparameters are identified, the final models are retrained using the full training interval to maximise the information available for learning before evaluation on independent test data. Automated hyperparameter tuning frameworks such as Optuna \cite{optuna}, together with validation monitoring and early stopping, are widely used and represent established best practice in modern machine learning workflows.

The final hyperparameters for the base and turbulence-aware models forecasting with a 60 minute horizon obtained from tuning are shown in Table \ref{tab:hyperparams}. The learning rate $\eta$ sets the shrinkage applied to each tree's contribution; smaller values slow the learning, requiring more trees but typically improving generalisation. The maximum depth limits how many splits each tree can make, bounding the order of feature interactions the model can represent and acting as a primary control on model complexity. Subsample is the fraction of training rows drawn at random for each tree, while column sample by tree is the fraction of features made available to each tree; both introduce stochasticity that reduces overfitting and decorrelates the ensemble. The minimum child weight sets the smallest total instance weight permitted in a leaf, so that larger values prevent the model from forming leaves that fit only a few samples. The parameter $\gamma$ is the minimum loss reduction required to make a further split, so higher values yield more conservative, less complex trees. Finally, $\gamma$ is the L2 regularisation term on the leaf weights, penalising large leaf values to further discourage overfitting.

\begin{table}
    \centering
    \caption{Tuned 60-minute Forecast Horizon Model Hyperparameters}
    \begin{tabular}{ccc}
        \textbf{Parameter} & \textbf{Base model} & \textbf{Turbulence-aware model} \\ \hline
        $\eta$ & 0.0182 & 0.0118 \\
        Max. depth & 4 & 5 \\
        Subsample & 0.423 & 0.373 \\
        Column sample by tree & 0.830 & 0.956 \\
        Min. child weight & 0.152 & 6.48 \\
        $\gamma$ & 3.66 & 0.140 \\
        $\lambda$ & 0.492 & 0.669 \\
    \end{tabular}
    \label{tab:hyperparams}
\end{table}

\subsection{Model evaluation and explainability}

Model performance is evaluated on the test interval using RMSE, MAE, $R^2$, and Pearson correlation. A persistence forecast $AE(t)$ is used as a baseline, and a persistence skill score is computed to quantify improvement,

\begin{equation}
    S = 1 - \frac{\mathrm{MSE}_{\mathrm{model}}}{\mathrm{MSE}_{\mathrm{persist}}},
\end{equation}
where $S>0$ indicates improvement over persistence.

Model behaviour is interpreted using Shapley Additive Explanations (SHAP) \cite{lundberg2017}. Global feature importance is quantified using mean absolute SHAP values, while dependence plots are used to examine nonlinear responses. Because solar wind features are strongly correlated, a correlation-aware SHAP formulation is employed \cite{aas2020}, preserving empirical feature dependencies and yielding physically meaningful attributions.

SHAP values are computed on a stratified subset of the test data to ensure representation across geomagnetic activity levels. Because the AE distribution is heavily weighted towards quiet time values, uniform random sampling of the test set would yield few high activity intervals and provide poor coverage of the large AE regime that is of greatest interest. To address this, the test set is partitioned into bins defined by quantiles of the observed AE distribution, with finer bins at higher activity levels, and a fixed number of samples is drawn at random without replacement from each bin. The quantile ranges with their sample numbers are as follows: Q0--50 with 1500 minutes, Q50--60 with 1000 minutes, Q60--80 with 800 minutes, Q80--90 with 500 minutes, and Q90--100 with 300 minutes. This over represents active intervals relative to their natural frequency while retaining a substantial quiet time population, so that the resulting attributions are not dominated by low activity samples and the model's behaviour during disturbed conditions is adequately characterised.

\subsection{Cost/loss analysis}

To evaluate whether the turbulence-aware forecast adds forecast value in comparison to the mean-values forecast, the application to operational space weather forecasting must be considered. This is carried out using cost/loss analysis as described in \citeA{richardson2011} and \citeA{owens2017}, taking into account different thresholds of AE index at which mitigating action would be taken.

An event is defined in relation to a threshold $T$,

\begin{equation}
E_t =
\begin{cases}
1, & AE_{\mathrm{obs}}(t) \ge T, \\
0, & \text{otherwise}.
\end{cases}
\end{equation}
As the forecast is deterministic (has no probability associated with a prediction), the probability of exceeding a threshold is

\begin{equation}
p_t =
\begin{cases}
1, & \hat{AE}(t) \ge T, \\
0, & \text{otherwise}.
\end{cases}
\end{equation}
The event rate for climatology is used to compare the benefit added by a model. In this case, this is taken as the long-term fraction of time that the chosen AE threshold is exceeded,

\begin{equation}
o = \mathbb{P}(E_t = 1)
  = \frac{1}{N} \sum_{t=1}^{N} E_t.
\end{equation}

For a user with a cost/loss ratio $r = C/L \in [0, 1]$, where $C$ is the cost of mitigating for an event and $L$ is the loss associated with an event, they will take mitigating action ($A_t = 1$) when $p_t = 1$ and not ($A_t$ = 0) when $p_t = 0$. This results in expenses associated with the forecast, climatology, and a perfect prediction,

\begin{equation}
E_{\mathrm{fcst}}
= C \sum_{t=1}^{N} A_t
+ L \sum_{t=1}^{N} E_t (1 - A_t),
\end{equation}

\begin{equation}
E_{\mathrm{clim}} =
\begin{cases}
C N, & o \ge r, \\
L \sum_{t=1}^{N} E_t, & o < r,
\end{cases}
\end{equation}

\begin{equation}
E_{\mathrm{perf}}
= C \sum_{t=1}^{N} E_t.
\end{equation}
The potential economic value of the forecast can then be found from these,

\begin{equation}
V
= \frac{E_{\mathrm{clim}} - E_{\mathrm{fcst}}}
       {E_{\mathrm{clim}} - E_{\mathrm{perf}}},
\end{equation}
which only depends on the ratio of cost and loss $r$. This allows for comparison of the potential economic value between different forecast users. $V = 1$ indicates a perfect forecast, $V = 0$ indicates no improvement over climatology, and $V < 0$ indicates that a forecast is worse than climatology when considering economic benefit.

\section{AE Prediction Performance} \label{sec:performance}

\begin{figure}
    \centering
    \includegraphics[width=\linewidth]{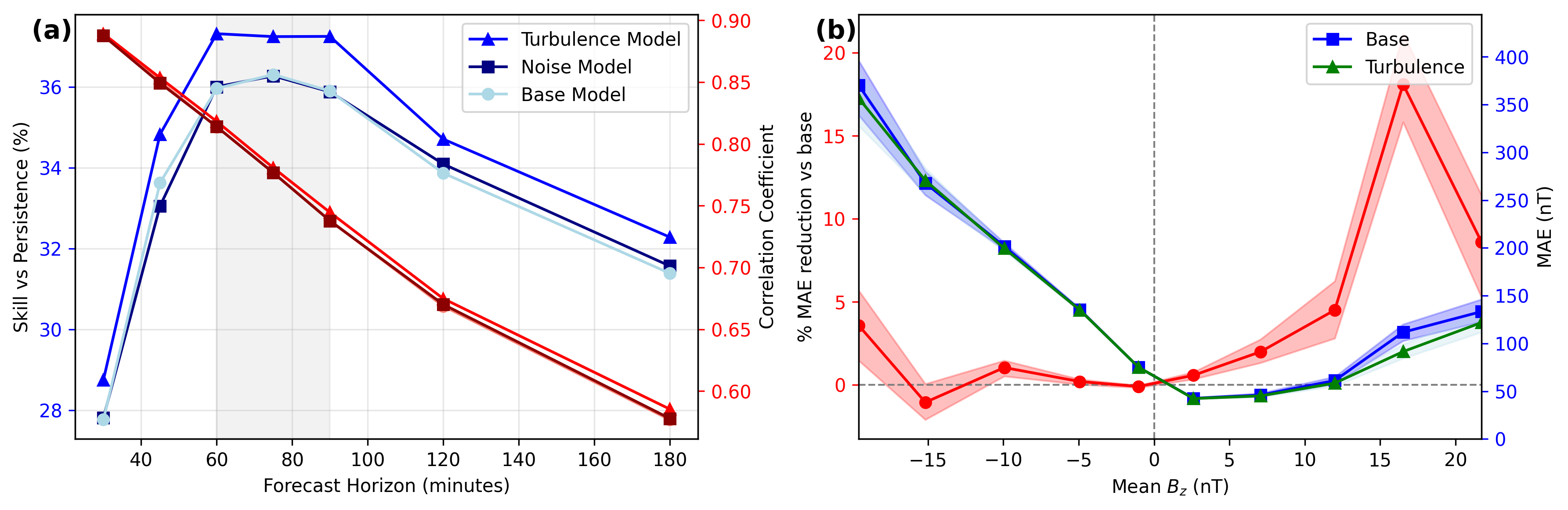}
    \caption{(a) Comparison of the performance of each of the base, turbulence and noise models for each of the chosen forecast horizons, with skill vs persistence shown in blue and correlation coefficient in red, with triangular markers for the turbulence-aware model, squares for the noise model, and circles for the base model. The range of horizons giving high quality predictions of AE index are shown by a grey shaded region. (b) Comparison of the mean absolute error (MAE) between the base (blue squares) and turbulence (green triangles) models for different values of $B_z$ averaged over the previous hour, with the binned percentage reduction by the turbulence model shown in red along with the range of values for each bin.}
    \label{fig:skill}
\end{figure}

A commonly used performance metric for assessing prediction of AE index is the correlation coefficient between the predicted and observed AE index. However, to compare between different target lags, this will almost always show a peak at the shortest lag due to information decay. Instead, the correlation is considered along with skill vs persistence, as shown in Figure \ref{fig:skill}a. For the base model, the skill peaks at a 75 minute lag window. In contrast, for the turbulence-aware model, the skill is sustained across multiple forecast horizons and is approximately equal for forecast horizons of 60, 75 and 90 minutes, indicating improved robustness to increasing lead time. As in both models the correlation is greatest at the shorter horizon, the models predicting 60 minutes into the future are chosen for further analysis. It is worth noting that the model replacing turbulence parameters with noise performs approximately equally to the base model, demonstrating that the improvement arises from physically meaningful structure in the turbulence parameters rather than simply increasing the number of inputs. Overall at 60 minutes, the correlation coefficient exceeds 0.8, only slightly lower than values reported for more complex neural network methods.

While short-term AE persistence contributes substantially to forecast skill, the turbulence-aware model provides a consistent and measurable improvement over both persistence and the base model. This demonstrates that the model is exploiting structured information in the upstream solar wind beyond simple autoregressive behaviour, and that turbulence descriptors provide additional predictive power for short timescale geomagnetic activity.

Motivated by the different effects of solar wind driving under southward and northward IMF, the improvement made by the turbulence-aware model is considered for different values of $B_z$ in GSM coordinates (Figure \ref{fig:skill}b). For southward $B_z$, there is little change to the mean absolute error between the two models, except for a small improvement at very large southward IMF. However, under northward IMF, the greater the magnitude of $B_z$, the greater the percentage improvement in MAE obtained by the turbulence model. This suggests that turbulence plays a greater role in modulating the effect of solar wind driving on AE index under northward IMF.

\begin{figure}
    \centering
    \includegraphics[width=\linewidth]{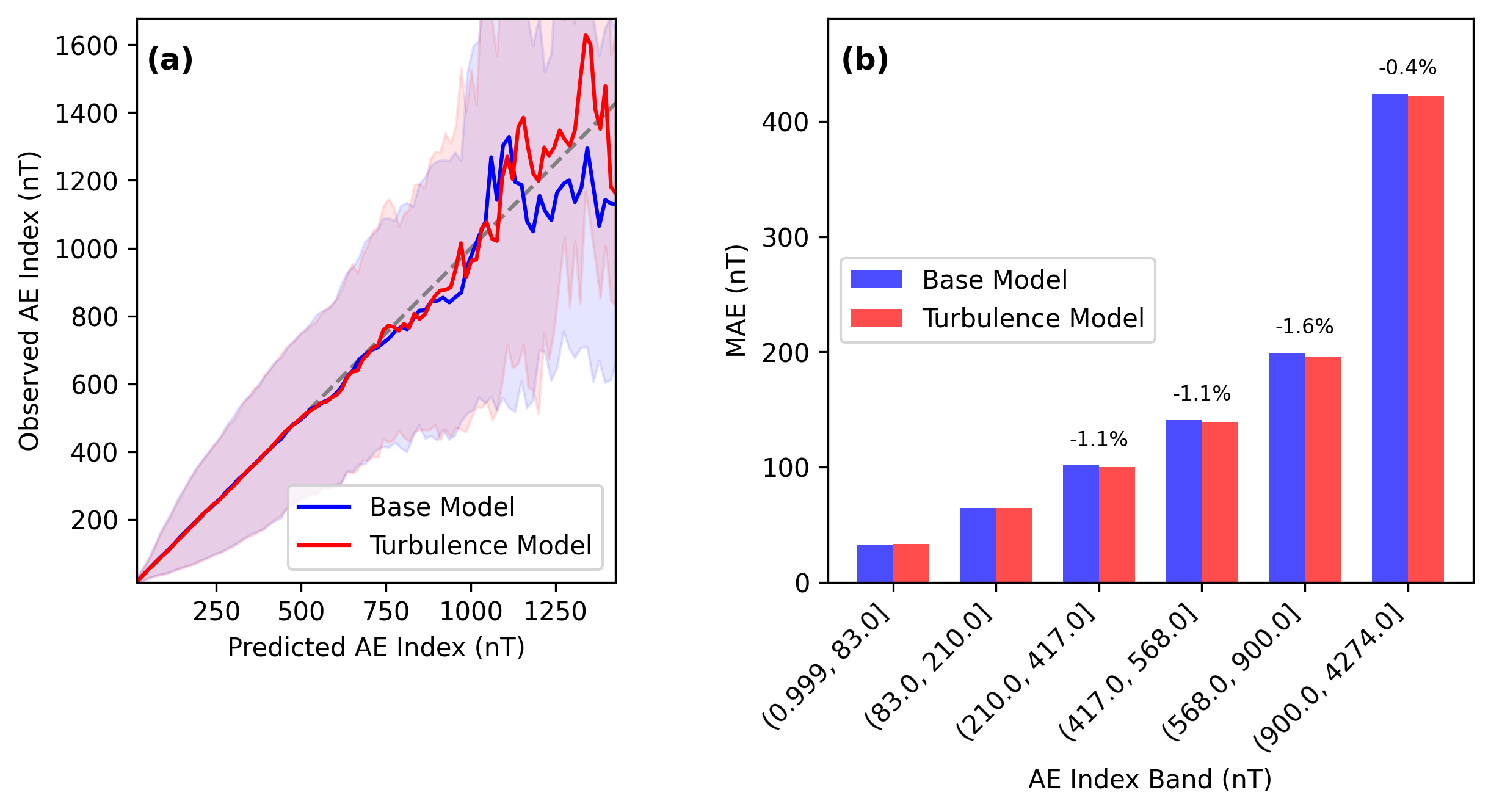}
    \caption{(a) Observed vs predicted AE index by the base model (blue) and turbulence model (red), binned in predicted AE index, with $AE_\text{obs} = AE_\text{pred}$ shown as a grey dashed line. The associated 10th and 90th percentiles are shown for each model as a confidence interval. (b) Average MAE for both the base (blue) and turbulence (red) models for bands of AE index, defined by the 0th, 50th, 75th, 90th, 95th, 99th and 100th percentiles as their edges.}
    \label{fig:performance}
\end{figure}

The improvement provided by the turbulence-aware model can also be seen to vary with the magnitude of AE index (Figure \ref{fig:performance}a). When plotting the observed vs predicted AE index, the turbulence model is more successful at following the $AE_\text{obs} = AE_\text{pred}$ line. The base model falls below this line for the higher AE values, indicating that it is overpredicting for more extreme events. This could result in the turbulence-aware model reducing false alarm rates for extreme space weather events. In Figure \ref{fig:performance}b, this is shown further by binning the AE index into bands by percentile. For lower values of AE index, the mean absolute error is approximately the same for both models. However, in the higher bands (75th percentile and above), the MAE is reduced by the turbulence model, by over 1\% between the 75th and 99th percentiles, and by 0.4\% above the 99th percentile. This indicates that overall the turbulence-aware model can predict more extreme events effectively, up until those that are the most extreme, where it does not give as much of an improvement. This is expected, as the most extreme space weather events are likely to have additional physical effects at play which are not described adequately by the training data, as the test data contains the two largest spikes in AE of the whole data period.

\begin{figure}
    \centering
    \includegraphics[width=\linewidth]{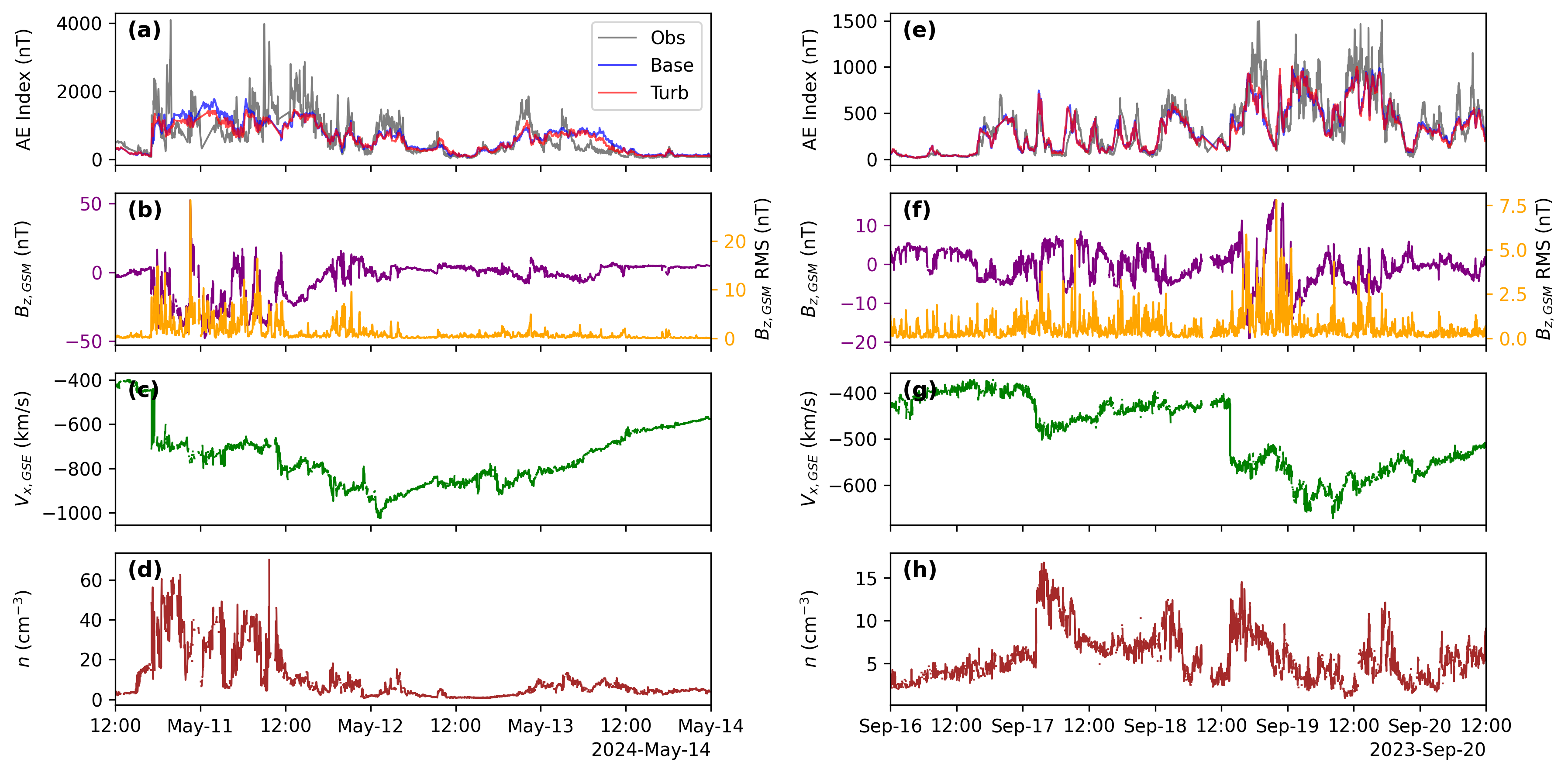}
    \caption{Two example space weather events with (a, e) observed AE index in grey, predicted (by the base model) AE index in blue, and predicted (by the turbulence model) AE index in red, (b, f) $z_{\text{GSM}}$ component of magnetic field $B_{z,\text{GSM}}$ in nT in purple and RMS fluctuations in this component $\delta B_{z,\text{rms}}$ in yellow, (c, g) $x_{\text{GSE}}$ component of solar wind velocity $V_{x,\text{GSE}}$ (km/s) in green, and (d, h) proton number density $n$ (cm$^{-3}$) in brown, for (a--d) May 11th to 14th 2024 and (e--h) 16th to 20th September 2023.}
    \label{fig:events}
\end{figure}

This can be seen by considering example space weather events. Figure \ref{fig:events} shows two events, with one being the extreme storm of May 2024 (Figure \ref{fig:events}a--d), and the other a more typical event from September 2023 (Figure \ref{fig:events}e--h). For May 2024, the timing of increases and decreases in AE index is predicted well (Figure \ref{fig:events}a). However, the magnitude of AE index is consistently underestimated. This is most likely due to the additional physical effects known to be occurring during this event due to multiple CME impacts \cite<e.g.>{tulasi2024, mactaggart2025} and additional flips in $B_y$ \cite{ohtani2025}, causing a strongly driven and compressed magnetopause \cite{beedle2025} and leading to additional ionospheric structure \cite<e.g.>{astafyeva2025, kuai2025}. The multiple impacts can be seen in $B_z$ in Figure \ref{fig:events}b, often with large $\delta B_{z, rms}$ associated with them, along with solar wind speed increases (Figure \ref{fig:events}c) and density enhancements (Figure \ref{fig:events}d).

September 2023 provides a view of a much more `typical' event. Figure \ref{fig:events}e shows that the predicted AE index corresponds much more closely in magnitude here, along with the timing of increases in decreases being predicted well. This event had a mixture of northward and southward IMF, with high $\delta B_{z, rms}$ corresponding to additional structures within the solar wind (Figure \ref{fig:events}f). This event also exhibits slow wind until around 12:00 on September 18th, before a sharp increase in wind speed with the impact of a structure (Figure \ref{fig:events}g), and shows density enhancements corresponding to these structures (Figure \ref{fig:events}h). Overall this demonstrates that the prediction works well for a more commonly occurring space weather event.

As previously discussed, comparing absolute error in prediction between events does not give the full picture due to their relative magnitudes in AE index. However, for these two events, we instead normalise by the median AE index to compare the root mean squared error. For May 2024, this median is 406 nT, and for September 2023, this is 302 nT. In May 2024, the base model normalised RMSE is 0.99, and the turbulence-aware model normalised RMSE is 0.94, resulting in a reduction of 0.05 times the median AE index. However, in September 2023, the base normalised RMSE is 0.52, and the turbulence-aware model normalised RMSE is 0.52, staying largely the same. This corresponds with it being forecasts for the most extreme events which are improved by using a turbulence-aware model, as discussed in Figure \ref{fig:performance}.

Overall, explicitly including turbulence parameters within the upstream solar wind dataset to predict AE index can improve the prediction skill and reduce errors in prediction of extreme space weather events. The percentage error reduction is greatest under northward IMF, suggesting that turbulence influences the effects of solar wind driving on AE index under northward IMF more strongly than under southward IMF. Although the percentage reduction in error is modest, the improvement is systematic, physically interpretable, and concentrated in regimes where baseline models perform worst --- namely under northward IMF and during elevated AE activity --- and therefore has disproportionate value for both scientific understanding and operational decision making.

\section{Parameter Dependence} \label{sec:parameters}

\subsection{Average Contributions}

\begin{figure}
    \centering
    \includegraphics[width=\linewidth]{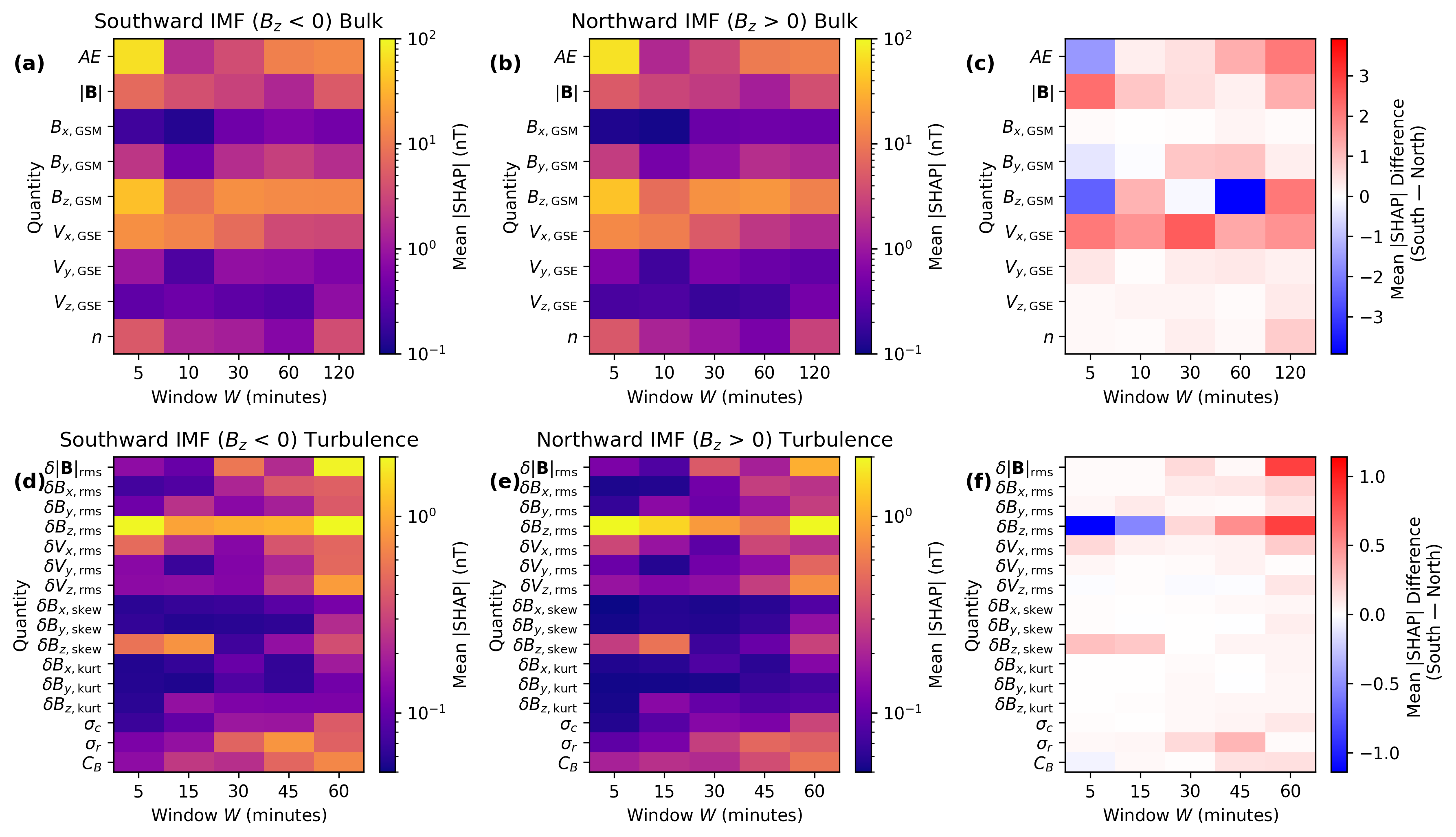}
    \caption{(a, b, d, e) Mean of the magnitudes of the SHAP values in the turbulence-aware model for each of the features shown as a heatmap, organised by physical variable and window $W$, for (a) bulk parameters under southward IMF, (b) bulk parameters under northward IMF, (d) turbulence parameters under southward IMF, and (e) turbulence parameters under northward IMF. (c) shows the difference in these values (southward -- northward) for the bulk parameters, and (f) for the turbulence parameters.}
    \label{fig:importance}
\end{figure}

Correlation-aware SHAP is used to quantify the mean contribution of each input variable to the turbulence-aware AE prediction. Contributions are separated by IMF orientation (Figure \ref{fig:importance}), reflecting differences in model behaviour under northward and southward conditions.

For bulk parameters (Figures \ref{fig:importance}a, b), the dominant predictor is the 5-minute mean AE index, capturing short-term persistence of auroral electrojet activity \cite{akasofu1964, mcpherron1970}. Intermediate windows (10--30 minutes) add little additional information, while contributions increase again at 60--120 minutes, indicating sensitivity to longer timescale clustering of geomagnetic activity \cite{tanskanen2009, pulkkinen2006}. The next most important parameter is the 5-minute mean $B_{z,\text{GSM}}$, consistent with its role in controlling dayside reconnection and energy input \cite{dungey1961, gonzalez1994}. Solar wind speed $V_{x,\text{GSE}}$ contributes primarily at short lags, with diminishing importance at longer times due to autocorrelation and reduced relevance for longer term prediction. Other bulk parameters, including $|\mathbf{B}|$ and $n$, provide secondary contributions, while $B_x$, $B_y$, and transverse velocity components have minimal impact.

The IMF dependent difference (Figure \ref{fig:importance}c) show a clear separation in controlling mechanisms. Under northward IMF, AE persistence and short timescale $B_z$ variations are more important, indicating a more gradual and variability-driven response. Under southward IMF, solar wind speed and $|\mathbf{B}|$ become more influential, consistent with stronger driving and enhanced coupling efficiency. Longer timescale AE history (120 minutes) is also more important under southward IMF, reflecting sustained or recurrent activity during prolonged driving.

For turbulence parameters, the dominant contributions arise from magnetic field fluctuations and their higher order structure (Figures \ref{fig:importance}d, e). RMS fluctuations in $B_{z,\text{GSM}}$ are the most important turbulence predictors across multiple timescales, with strong contributions at both short (5 min) and long (60 min) windows. This indicates sensitivity to both rapid IMF variability and longer timescale fluctuations controlling coupling efficiency. RMS fluctuations in $|\mathbf{B}|$ at 60 minutes are similarly important, highlighting the role of turbulent magnetic energy. Higher order statistics also contribute, with skewness of $B_z$ fluctuations important at short timescales, consistent with the impact of intermittent southward excursions.

At longer timescales, parameters describing turbulence structure become more important. Residual energy $\sigma_r$ contributes for windows $\geq$ 30 minutes, indicating sensitivity to magnetic--kinetic energy imbalance, while compressibility $C_B$ becomes important at 45--60 minutes, suggesting that fluctuation geometry influences coupling. Velocity fluctuations play a secondary role, with only weak contributions at long timescales.

The IMF dependence of turbulence contributions is systematic. Short timescale $B_z$ fluctuations are more important under northward IMF, indicating enhanced sensitivity to directional variability when coupling is weak. In contrast, longer timescale fluctuations, magnetic energy content, and structural parameters ($\sigma_r$, $C_B$) are more important under southward IMF, reflecting cumulative turbulent driving under strong coupling. Overall, these results show that turbulence influences AE through different mechanisms depending on IMF orientation: short timescale variability dominates under weak driving, while longer timescale amplitude and structure control the response under strong driving.

\subsection{Behaviour of Bulk Features}

\begin{figure}
    \centering
    \includegraphics[width=\linewidth]{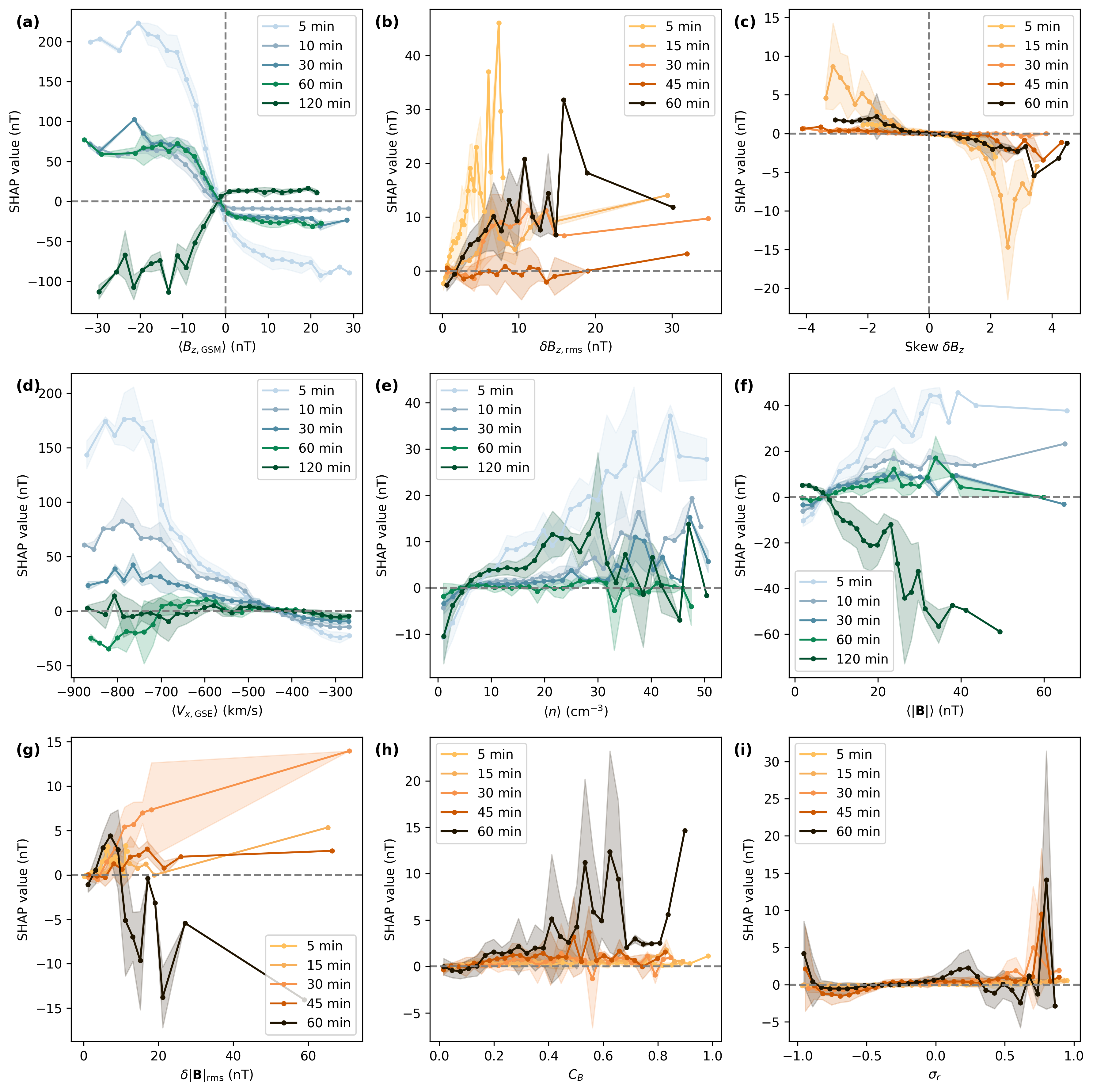}
    \caption{Median values of SHAP value for bins of (a) $z$ component of magnetic field $B_{z,\text{GSM}}$, (b) RMS fluctuations of $B_z$ $\delta B_{z,\text{rms}}$, (c) skew of $\delta B_z$, (d) $x$ component of velocity $V_{x,\text{GSE}}$, (e) proton number density $n$, (f) magnetic field strength $|\mathbf{B}|$, (g) RMS fluctuations of field strength $\delta |\mathbf{B}|_{\text{rms}}$, (h) magnetic compressibility $C_B$, and (i) residual energy $\sigma_r$. Bulk parameters (a, d, e, f) are shown windowed over 5, 10, 30, 60 and 120 minutes, and turbulence parameters (b, c, g, h, i) are shown windowed over 5, 15, 30, 45 and 60 minutes. The median values $\pm\sigma$ are shown.}
    \label{fig:shap_sum}
\end{figure}

Figure \ref{fig:shap_sum}a, d, e, f shows the SHAP dependence of the most influential bulk solar wind and IMF parameters. At short lead times (5--60 min), the model response is dominated by IMF $B_z$, solar wind speed $V_x$, and magnetic field magnitude $|\mathbf{B}|$, consistent with established solar wind--magnetosphere coupling physics \cite{newell2007, borovsky2003, luo2013}. At longer lead times (120 min), these dependencies weaken or change sign, reflecting the increasing influence of internal magnetospheric dynamics and the reduced direct control of upstream conditions \cite{weigel2003}.

IMF $B_z$ (Figure \ref{fig:shap_sum}a) shows the expected strong sensitivity to southward conditions at short lead times, with increasingly negative $B_z$ producing large positive contributions to AE. For strong southward IMF ($B_z \lesssim$ -10 nT), the response saturates, indicating that additional increases in driving do not produce proportionally larger AE. This plateau is consistent with known nonlinear limits in magnetospheric coupling, including ionospheric conductance constraints and feedback processes \cite{siscoe2002, borovsky2021b, myllys2016}. As lead time increases, the magnitude of the $B_z$ dependence decreases, and by 120 minutes the relationship reverses, with strong southward IMF associated with reduced AE. This reflects the transition from directly driven behaviour to phase-dependent evolution, where strong driving is followed by recovery or unloading \cite{weigel2003, borovsky2021b}.

Solar wind speed $V_x$ (Figure \ref{fig:shap_sum}d) exhibits a monotonic dependence at short lead times, with faster flows producing larger positive contributions, consistent with enhanced convective electric field driving \cite{newell2007, luo2013}. This dependence weakens steadily with increasing lead time and is negligible by 120 minutes, indicating that solar wind speed primarily controls short-term energy input rather than longer term AE evolution.

Proton density $n$ (Figure \ref{fig:shap_sum}e) has a weaker and more transient influence. At short lead times, increasing density produces positive contributions, consistent with enhanced dynamic pressure and magnetopause compression. However, this dependence collapses at longer lead times and becomes non-monotonic, indicating limited predictive value beyond short timescales \cite{weigel2003, borovsky2021b}.

Magnetic field magnitude $|\mathbf{B}|$ (Figure \ref{fig:shap_sum}f) shows a threshold-like response at short lead times, with low values associated with reduced AE and a transition to positive contributions above $\sim$5-10 nT. At higher field strengths, the response flattens, indicating diminishing sensitivity under strong driving. As with $B_z$, the dependence weakens and reverses at longer lead times, reflecting the reduced predictive role of instantaneous upstream conditions and the increasing importance of internal dynamics.

Overall, these results show that AE predictability at short lead times is controlled by IMF orientation and solar wind driving in a manner consistent with established coupling physics. At longer lead times, the influence of these drivers diminishes and, in some cases, reverses, highlighting the transition to internally governed magnetospheric dynamics and setting a fundamental limit on solar wind-driven predictability.

\subsection{Behaviour of Turbulence Features}

Figure \ref{fig:shap_sum}b, c, g, h, i shows the SHAP dependence of turbulence parameters, demonstrating how fluctuation amplitude, intermittency, and structure influence the predicted AE index. The dominant turbulence signal arises from magnetic field variability. RMS fluctuations in $B_z$ exhibit a strong positive contribution at short timescales (5 min), with continued but weaker influence at longer windows (15--60 min). This indicates that both rapid directional variability and longer timescale fluctuations in IMF orientation enhance geomagnetic response, consistent with the central role of $B_z$ in controlling coupling.

In contrast, RMS fluctuations in the total field $|\mathbf{B}|$ show a weaker and more scale-dependent effect. While intermediate timescales ($\sim$30 min) show a modest positive contribution, large amplitude fluctuations at longer timescales (60 min) produce negative SHAP values, indicating that strong magnetic variability does not necessarily translate into increased geomagnetic activity. This suggests a regime in which increased fluctuation amplitude reflects less efficient or more disorganised coupling rather than stronger driving.

Higher order and structural turbulence parameters exhibit more selective behaviour. The skewness of $B_z$ fluctuations is most influential at intermediate timescales ($\sim$15 min), with negative skewness (intermittent southward excursions) producing positive contributions and positive skewness producing negative contributions. This highlights the importance of intermittent, asymmetric fluctuations rather than purely Gaussian variability. At shorter and longer windows, the influence of skewness is weak, indicating that intermittency is most effectively captured at intermediate timescales.

Parameters describing turbulence structure and energy partition become important at longer timescales. Residual energy $\sigma_r$ shows a strong positive contribution as it approaches unity, indicating sensitivity to magnetic--kinetic imbalance in the fluctuations. Compressibility $C_B$ also becomes important at 45--60 min, with peak contributions at moderate values ($\sim$0.5) and reduced influence at very high compressibility, suggesting that moderately structured turbulence is most effective at enhancing geomagnetic response.

Overall, these results show that turbulence contributes to AE prediction through distinct physical pathways across scales. Short timescale variability in $B_z$ enhances immediate coupling, intermittency at intermediate scales modulates the occurrence of geoeffective fluctuations, and longer timescale properties such as energy partition and fluctuation geometry influence the cumulative efficiency of energy transfer. Importantly, these effects persist even when mean IMF conditions are included, indicating that turbulence provides complementary predictive information beyond large scale solar wind driving.

\section{Model Benefits (Cost/Loss Analysis)} \label{sec:impact}

\begin{figure}
    \centering
    \includegraphics[width=\linewidth]{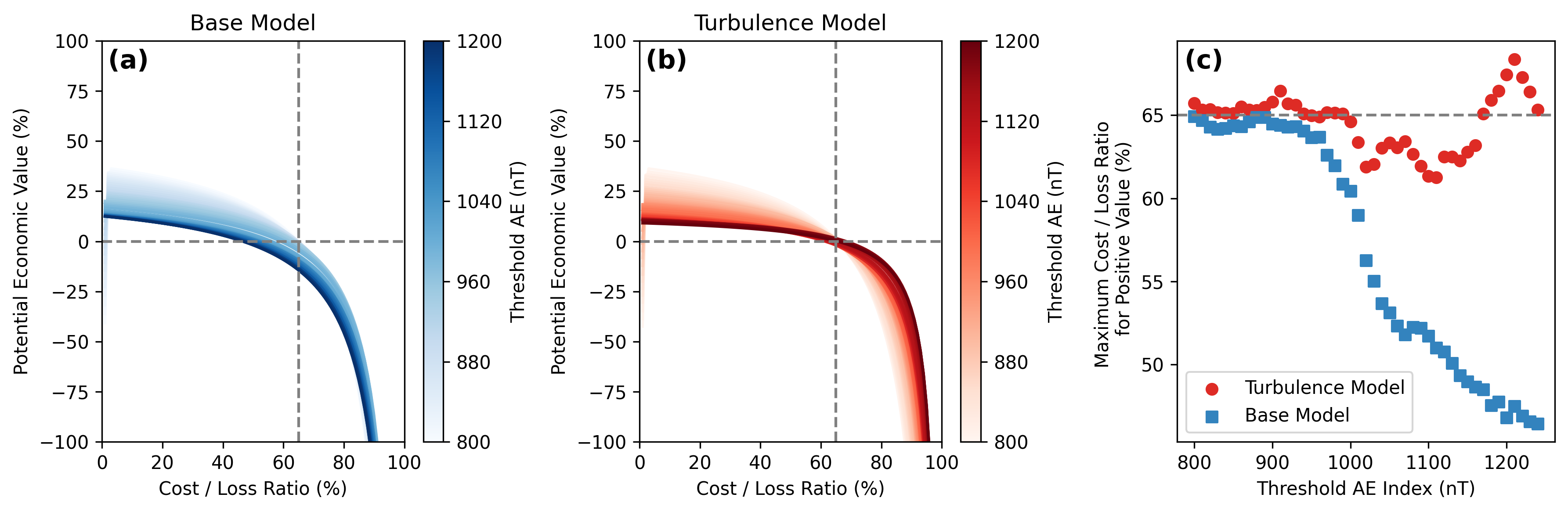}
    \caption{Potential economic value $V$ against cost/loss ratio $r$ for the base model in (a) and turbulence model in (b), for a range of thresholds of AE between 800 and 1200 nT. (c) The $x$-intercept of each of the curves plotted against the threshold AE index for the turbulence model (red circles) and the base model (blue squares).}
    \label{fig:cl_curves}
\end{figure}

Figures \ref{fig:cl_curves}a and b present the potential economic value of the AE forecasts as a function of cost-loss ratio for a range of AE index thresholds. Following the cost-loss framework defined in Section \ref{sec:methods}, the potential economic value $V$ is shown as a percentage, where 100\% denotes the value of a perfect forecast and 0\% denotes no improvement over climatology. Negative values indicate a forecast less useful than climatology. Recall that $V$ depends only on the user's cost-loss ratio $r = C/L$, so each curve describes how the forecast's value varies across users with different relative costs of taking mitigating action. The cost-loss ratio at which a curve crosses zero therefore marks the boundary beyond which climatology is the better basis for action, and the span of $r$ over which the curve is positive indicates the breath of users for whom the forecast is beneficially beneficial.

For the base model, the potential economic value systematically decreases as the AE threshold increases, with the value curves shifting downward and the range of cost-loss ratios for which the forecast provides positive value narrowing. In particular, the cost-loss ratio at which the potential value falls to zero decreases monotonically with increasing threshold, indicating that the base model becomes progressively less useful for decision making as the focus shifts to more extreme AE events. This behaviour reflects a deterioration in the model's ability to provide reliable discrimination between events and non-events at higher thresholds, consistent with increasing false alarms or missed events under more stringent conditions.

In contrast, the turbulence-augmented model exhibits markedly different behaviour. While the overall magnitude of the potential value varies with threshold, the cost-loss ratio at which the value drops to zero remains approximately constant across all AE thresholds considered. This indicates that the turbulence model retains economic usefulness over a similar range of decision maker preferences, even as the event definition becomes more extreme. Figure \ref{fig:cl_curves}c highlights this distinction: the zero-value intercept for the turbulence model is largely invariant, whereas the intercept for the base model decreases steadily with increasing threshold. These results demonstrate that incorporating turbulence parameters stabilises forecast value across event severities, implying improved robustness of the forecast for high impact, low probability events. From a decision making perspective, this suggests that the turbulence model maintains practical utility for mitigation strategies targeting extreme geomagnetic activity, whereas the base model's usefulness rapidly degrades as event severity increases.

\section{Discussion} \label{sec:discussion}

This study shows that explicitly incorporating turbulence-related descriptors into a data driven AE forecasting model produces systematic and physically interpretable improvements over a baseline model that uses only mean solar wind and IMF parameters. Although the absolute reduction in global error metrics is modest, the improvements are concentrated in regimes where baseline performance is weakest, including periods of northward IMF and elevated AE activity. Importantly, these gains extend beyond average error reduction to operationally relevant measures, such as reduced false alarms and improve cost-loss performance for high impact events.

Interpretation of both bulk and turbulence features using SHAP is broadly consistent with established solar wind--magnetosphere coupling physics. Short-lead forecasts are dominated by IMF orientation, solar wind speed, and magnetic field strength, while turbulence descriptors provide additional information related to variability, intermittency, and Alfv\'enic structure that is not captured by mean parameters alone. At longer lead times, the reduced influence of upstream drivers and the emergence of sign reversals reflect increasing control by internal magnetospheric dynamics rather than deficiencies in the model.

The improved performance of the turbulence-aware model supports the view that solar wind fluctuations play an active role in regulating magnetospheric energy transfer, rather than acting solely as noise around a mean driving state. Turbulence parameters encode information about the structure and coherence of the upstream plasma, which can influence reconnection efficiency, plasma transport, and the timing of energy release in the magnetosphere.

Previous observational studies have shown that Alfv\'enic turbulence can enhance auroral activity even under weak mean driving, while intermittent fluctuations may locally increase reconnection rates. The present results are consistent with this picture: turbulence descriptors retain predictive influence even when mean IMF strength and orientation are included, indicating that they encode complementary information rather than simply acting as proxies for stronger large scale driving. The fact that improvements are most pronounced under northward IMF further supports the interpretation that variability becomes particularly important when steady reconnection is suppressed.

Forecast improvements are strongly regime dependent. At short lead times, turbulence-aware forecasts better capture rapid AE enhancements and elevated activity levels, while at longer lead times the additional information becomes less influential as internal magnetospheric dynamics dominate. This behaviour is consistent with previous work showing that AE predictability based on upstream solar wind measurements is fundamentally limited at multi-hour timescales.

The plateauing or sign reversal of some SHAP dependencies at longer horizons should therefore not be interpreted as unphysical behaviour, but rather as a reflection of phase-dependent statistics in a nonlinear, history dependent system. Strong upstream driving is often associated with earlier AE enhancement; at later times, the conditional expectation may favour declining activity as the system unloads stored energy. These results reinforce the existence of an effective upper bound on AE predictability using near-Earth solar wind measurements alone. Turbulence information can move performance closer to this bound but cannot eliminate it.

The SHAP-based interpretation of bulk solar wind and IMF parameters aligns with established coupling physics. The strong short lead sensitivity to southward IMF reflects reconnection-controlled energy transfer, while the plateauing response under strong southward IMF is consistent with nonlinear saturation effects related to conductance limitations and magnetosphere--ionosphere feedbacks. The dominant influence of solar wind speed and magnetic field strength at short lead times is similarly consistent with their role in settings the convective electric field.

At longer lead times, the weakening and occasional sign reversal of these dependencies reflects the increasing importance of internal dynamics and event timing. These effects highlight the limitations of marginal feature interpretations in an autocorrelated system and underscore the need to interpret SHAP results as statistical attributions rather than direct causal relationships.

From an operational perspective, the results demonstrate that modest improvements in average forecast error can translate into meaningful gains when evaluated using decision-relevant metrics. The turbulence-aware model consistently improves cost-loss performance, particularly for higher thresholds and cost-loss ratios, indicating reduced false alarms and improved detection of impactful events. These findings suggest that incorporating physically motivated descriptors of variability may be a productive strategy for improving space weather forecasts, complementing advances in machine learning methodology.

Several limitations should be noted. The analysis is restricted to near-Earth solar wind measurements, limiting lead time and excluding explicit information about the internal magnetospheric state. Incorporating proxies for magnetospheric preconditioning may further improve forecasts, particularly at longer horizons. In addition, the turbulence descriptors have been derived from finite-resolution measurements and may not fully capture physical processes occurring below this resolution; future work could investigate the effects of higher resolution data. This work also made use of relatively simple machine learning models in order to demonstrate the benefit of adding turbulence-derived parameters, however, prediction skill could be improved further by combining this with more sophisticated models.

Other future work could explore adaptive windowing strategies, integrate turbulence diagnostics with state-space models, or apply similar approaches to other geomagnetic indices. More broadly, combining physically interpretable turbulence measures with data driven forecasting frameworks offers a promising path toward improved performance and understanding in space weather prediction.

\section{Conclusions} \label{sec:conclusions}

This study has examined whether the inclusion of explicit solar wind turbulence descriptors improves short-term forecasts of the auroral electrojet (AE) index within a machine learning framework. Two XGBoost models were developed and evaluated: a baseline model using mean solar wind parameters, and a turbulence-aware model augmented with physically motivated measured of variability.

The turbulence-aware model consistently outperforms the baseline across a range of lead times and evaluation metrics. It also outperforms a noise surrogate model, showing that the improvement is not just due to a greater number of parameters. While improvements in global error statistics are modest, they are systematic and most pronounced in regimes where baseline forecasts perform poorly, including under northward IMF conditions and during elevated AE activity. Importantly, these gains translate into improved operational performance, as demonstrated by reduced false alarms and improved cost-loss outcomes for higher AE thresholds.

Analysis of feature attributions confirms that bulk solar wind parameters dominate short-lead AE predictability, while turbulence descriptors provide complementary information that enhances forecast robustness. At longer lead times, forecast skill degrades for both models, consistent with increasing influence of internal magnetospheric dynamics and known limits to predictability based on upstream solar wind measurements alone.

Overall, these results demonstrate that incorporating physically interpretable descriptors of solar wind variability can yield meaningful improvements in geomagnetic activity forecasts without increasing model complexity. This approach provides a practical pathway for enhancing space weather prediction systems and motivates further exploration of turbulence-aware, physics-informed data driven models for other geomagnetic indices.

\section*{Open Research Section}
The baseline, turbulence-aware and noise-surrogate AE prediction models, along with instructions on how to run them, can be found at \url{https://github.com/carawaters/TurbulenceSpaceWeatherForecast} \cite{Waters_TurbulenceSpaceWeatherForecast_2026}. We acknowledge use of NASA/GSFC's Space Physics Data Facility's OMNIWeb service, and OMNI data \cite{omni}. The AE index was obtained from the World Data Center for Geomagnetism, Kyoto (\url{https://wdc.kugi.kyoto-u.ac.jp/}), as detailed by \citeA{davis1966}. Data was accessed using PySPEDAS, an implementation of the Space Physics Environment Data Analysis Software (SPEDAS) framework in Python \cite{grimes2022} (\url{https://pyspedas.readthedocs.io/en/latest/}), version 2.0.6 released 14th January 2026. The XGBoost model was implemented using the XGBoost Python package \cite{xgboost} (\url{https://xgboost.readthedocs.io/en/stable/}), version 3.1.3 released 9th January 2026, and was optimised using Optuna \cite{optuna} (\url{https://github.com/optuna/optuna}), version 4.6.0 released 20th November 2025. Model interpretation used SHAP (SHapley Additive exPlanations) \cite{shap} (\url{https://shap.readthedocs.io/en/latest/index.html}), version 0.50.0 released 11th November 2025, with the Correlated SHAP implementation (\url{https://github.com/Fraunhofer-SCAI/corr_shap}), based on the method described by \citeA{aas2020}, version 0.0.2 released 14th November 2023.

\section*{Conflict of Interest declaration}
The authors declare there are no conflicts of interest for this manuscript.

\acknowledgments
CLW and CHKC are supported by UKRI Future Leaders Fellowship MR/W007657/1. CHKC is also supported by STFC Consolidated Grant ST/X000974/1. MJO is part-funded by STFC Grant ST/V000497/1 and NERC Grant NE/Y001052/1. This research utilised Queen Mary's Apocrita HPC facility, supported by QMUL Research-IT (\url{http://doi.org/10.5281/zenodo.438045}). The authors thank A. Smith for discussions about correlation-aware feature attribution.

\bibliography{references}

\end{document}